\documentclass{ws-procs961x669}            % default, citations in superscript
\begin{document}
\title{Information loss problem and roles of instantons\footnote{Proceedings for \textit{The 2nd LeCosPA Symposium: Everything about Gravity, Celebrating the Centenary of Einstein's General Relativity}. Talk on December 17, 2015, Taipei, Taiwan.}}

\author{Dong-han Yeom}

\address{Leung Center for Cosmology and Particle Astrophysics,\\
National Taiwan University, Taipei 10617, Taiwan\\
E-mail: innocent.yeom@gmail.com}

\begin{abstract}
In order to understand the information loss problem, non-perturbative effects will do significant roles. Instantons are in general helpful for this purpose. There are various and rich thin-shell instantons and tunneling channels that eventually result a trivial geometry without a singularity nor an event horizon. We further discuss that there are some subtle examples in asymptotic de Sitter cases that need further investigations.
\end{abstract}

%\keywords{Style file; \LaTeX; Proceedings; World Scientific Publishing.}

\bodymatter

\section{Introduction: effective loss of information}

\begin{figure}[h]
\begin{center}
\includegraphics[scale=0.7]{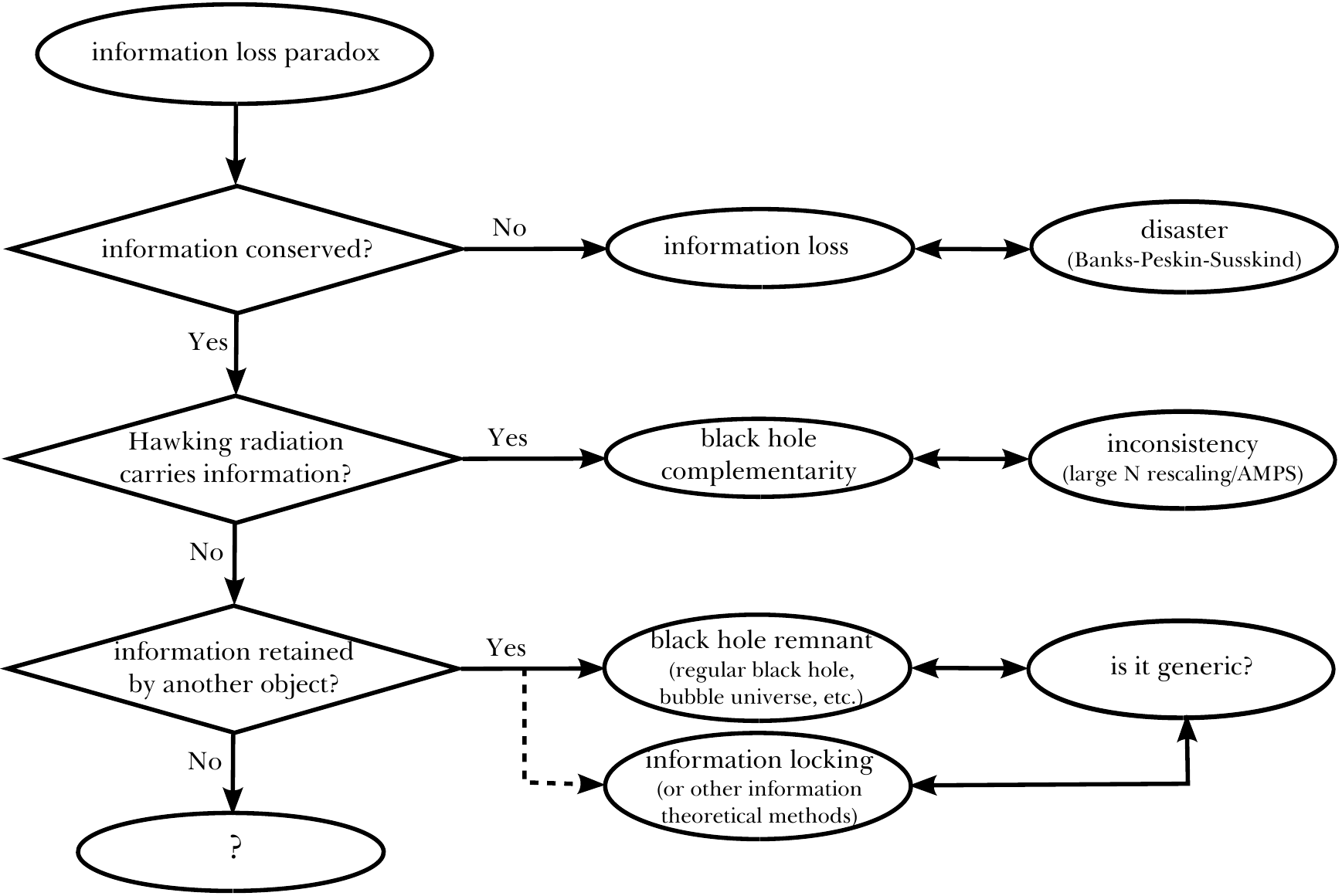}
\end{center}
\caption{\label{fig:flowchart}Flowchart on resolutions of the information loss problem.}
\end{figure}

In order to resolve the information loss problem \cite{Hawking:1976ra}, various ideas have been suggested in the literature (Fig.~\ref{fig:flowchart}) \cite{Chen:2014jwq}. However, it is also fair to say that various ideas have their own problems. If the black hole evolution is not unitary, then in itself it can cause a serious problem \cite{Banks:1983by}. If Hawking radiation contains information \cite{Susskind:1993if}, then we finally reach the inconsistency \cite{Yeom:2009zp}\cite{Almheiri:2012rt}. If information should be retained by another objects, the problem is that we still do not have a generic argument to realize this \cite{Chen:2014jwq}. Then what does this mean? If information should be conserved, but it is not attached by Hawking radiation nor by a remnant, then where is the information? One interesting way to explain this dilemma is to think that the wave function of the universe contains information; however, semi-classical observers, like us, cannot access the information since we should see a specific semi-classical geometry \cite{Sasaki:2014spa} (`question mark' in Fig.~\ref{fig:flowchart}). We name this idea as the \textit{effective loss of information}.

Then, how can we realize this idea? In order to do this, we need to understand the wave function of the universe \cite{DeWitt:1967yk}, while this is a genuinely quantum gravitational problem. Even though there is no consensus on quantum theory of gravity, the Euclidean path integral approach \cite{Hartle:1983ai} can be a highway to understand, at least, the skeleton of the wave function of the universe, by using instantons.

\section{Euclidean path integral approach}

According to the Euclidean path integral approach, the ground state wave function \cite{Hartle:1983ai} that satisfies the Wheeler-DeWitt equation \cite{DeWitt:1967yk} is presented by the following form:
\begin{eqnarray}
\langle f | i \rangle = \int_{i \rightarrow f_{j}} \mathcal{D} g_{\mu\nu} \mathcal{D}\phi \; e^{-S_{\mathrm{E}}\left[g_{\mu\nu},\phi \right]},
\end{eqnarray}
where $| i \rangle$ is the initial state, $| f \rangle = \sum_{j} a_{j} | f_{j}\rangle$ is the final state as a superposition of classical states $\{ |f_{j} \rangle \}$ with probability weights $\{ a_{i} \}$, $S_{\mathrm{E}}$ is the Euclidean action, and we sum over all geometries and field combinations that connect from $| i \rangle$ to a certain classical state $| f_{j} \rangle$.

Since this path integral is not very easy to do, we approximate this by summing over only on-shell solutions, or instantons, relying on the steepest-descent approximation. Then we can simplify
\begin{eqnarray}
\langle f | i \rangle \simeq \sum_{i \rightarrow f_{j}} e^{-S^{\textrm{on-shell}}_{\mathrm{E}}}.
\end{eqnarray}

If there exists a history with non-zero probability that connects from $| i \rangle$ to $| f_{k} \rangle$ and if this history has no event horizon nor singularity, i.e., if this has a \textit{trivial geometry}, then information will be conserved through the geometry \cite{Maldacena:2001kr}\cite{Hawking:2005kf}. Hence, in this sense, the wave function of the universe has information. However, since its probability can be highly suppressed due to the entropy cost of the instanton, perhaps the most probable history is the non-trivial geometry that is estimated by the perturbative quantum field theory in a curved spacetime; and in this sense, the semi-classical observer will lose information \cite{Sasaki:2014spa}.

Now the task is to find such an instanton that mediates a trivial geometry. Still we do not have a generic solution, but we can find various interesting examples using the thin-shell approximation.

\section{Examples of thin-shell instantons}

We consider a spacetime with the metric, $ds_{\pm}^{2}= - f_{\pm}(R) dT^{2} + f_{\pm}(R)^{-1} dR^{2} + R^{2} d\Omega^{2}$, where a thin-shell locates at $r$: outside the shell is $r < R$ (denoted by $+$) and inside the shell is $R < r$ (denoted by $-$). The thin-shell satisfies the metric $ds^{2} = - dt^{2} + r^{2}(t) d\Omega^{2}$. We assume that the outside and inside the shell look like $f_{\pm}(R) = 1 - 2M_{\pm}/R - R^{2}/\ell_{\pm}^{2}$, where $M_{+} > 0$ and $M_{-} = 0$ are the mass parameters of each region and $\ell^{2}_{\pm} = 3/8\pi U_{\pm}$ denotes the amount of the vacuum energy ($\ell_{\pm}^{2}$ can be chosen negative in order to present negative vacuum energy).

The equation of motion of the thin-shell is determined by the junction equation \cite{Israel:1966rt}: $\epsilon_{-} \sqrt{\dot{r}^{2}+f_{-}(r)} - \epsilon_{+} \sqrt{\dot{r}^{2}+f_{+}(r)} = 4\pi r \sigma$, where $\epsilon_{\pm} = \pm 1$ denotes the direction of the outward normal direction and $\sigma$ is a tension parameter which is a constant for a scalar field case, while in general this can be a function of $r$ and should satisfy physical constraints.

\subsection{Asymptotic anti-de Sitter and Minkowski}

For asymptotic anti-de Sitter or Minkowski cases, many models have been observed. For example,
\begin{itemize}
\item[$-$] \textit{Sasaki and Yeom} \cite{Sasaki:2014spa} considered true vacuum bubbles in anti-de Sitter background.
\item[$-$] \textit{Lee, Lee and Yeom} \cite{Lee:2015rwa} considered a magnetic shell in 3-dimensional anti-de Sitter background. Due to the magnetic field, there appeared much diverge tunneling channels.
\item[$-$] \textit{Chen, Dom\`enech, Sasaki and Yeom} \cite{Chen:2015lbp} considered vacuum bubbles with various tensions that can be in principle justified by physical matters. Some of these examples can show an initial condition that is free from the white hole initial singularity.
\item[$-$] \textit{Chen, Hu and Yeom} \cite{Chen:2015ibc} considered instanton creation from a no-shell combination to a two-shell combination. The probability can be well-defined, not only relying on the Euclidean method \cite{Gregory:2013hja}, but also from the Hamiltonian method \cite{Fischler:1990pk}.
\end{itemize}
These kind of examples should be generalized in the sense that \textit{(i) it should not rely on vacuum structures} and \textit{(ii) it should not rely on the thin-shell approximation}. We remain these topics as future research directions.

\begin{figure}
\begin{center}
\includegraphics[scale=0.28]{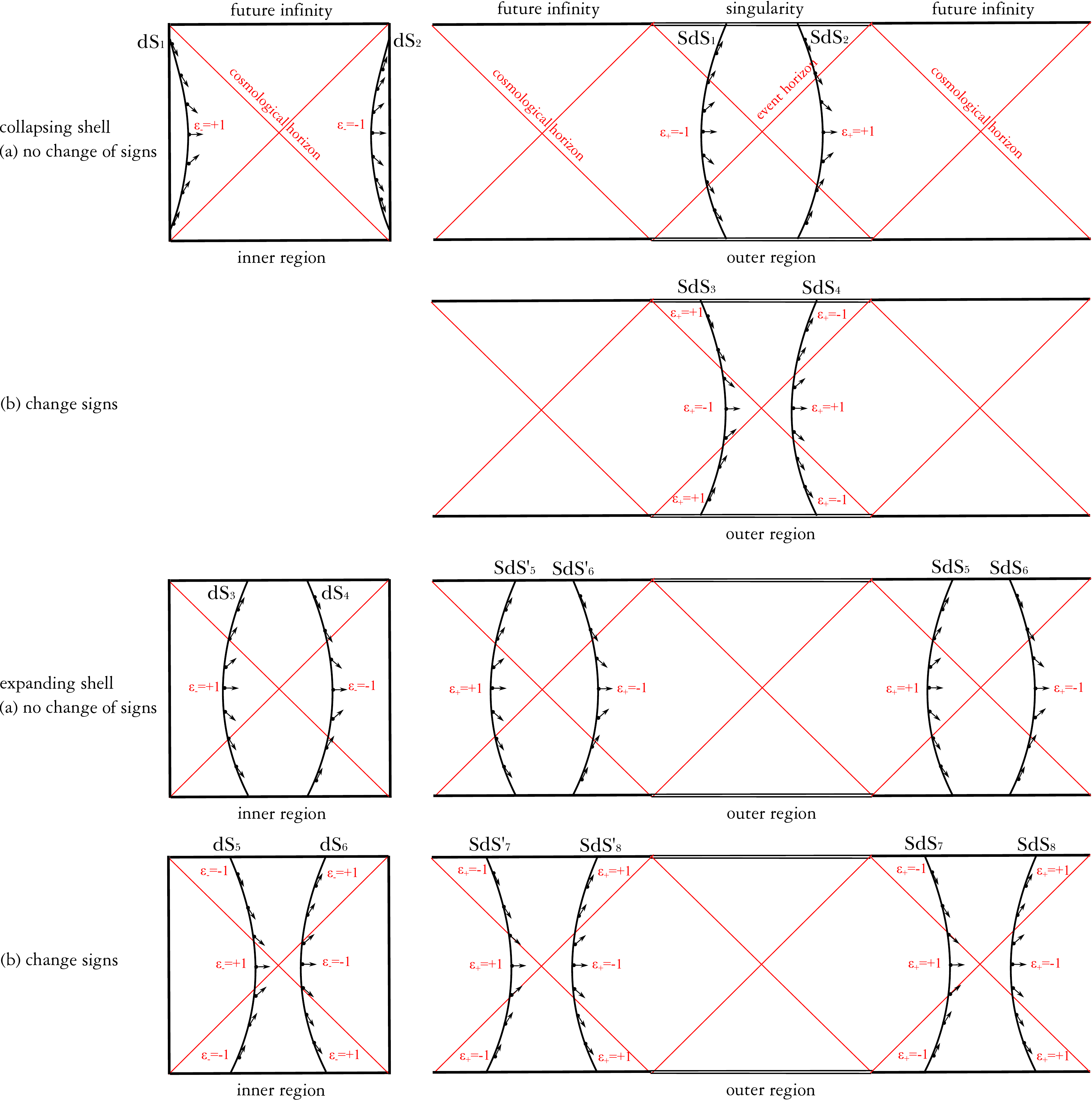}
\caption{\label{fig:signs_dS}Summary of the shell dynamics for symmetric solutions.}
\end{center}
\end{figure}

\subsection{Asymptotic de Sitter and subtle issues}

For asymptotic de Sitter cases, there appears further subtle issues. In this section, we see more details.

If the tension is a constant, then after simple calculations, we can reduce the junction equation by $\dot{r}^{2} + V(r) = 0$, where
\begin{eqnarray}
V(r) &=& f_{+}(r)- \frac{\left(f_{-}(r)-f_{+}(r)-16\pi^{2} \sigma^{2} r^{2}\right)^{2}}{64 \pi^{2} \sigma^{2} r^{2}}\\
&=& 1 - \left(\frac{1}{\ell_{+}^{2}} + \frac{\mathcal{B}^{2}}{64 \pi^{2} \sigma^{2}}\right) r^{2} - 2M_{+}\left( 1 + \frac{\mathcal{B}}{32 \pi^{2} \sigma^{2}} \right) \frac{1}{r}  - \frac{M_{+}^{2}}{16 \pi^{2} \sigma^{2}} \frac{1}{r^{4}}
\end{eqnarray}
and $\mathcal{B} \equiv - \ell_{-}^{-2} + \ell_{+}^{-2} - 16 \pi^{2} \sigma^{2}$.

\begin{figure}
\begin{center}
\includegraphics[scale=0.17]{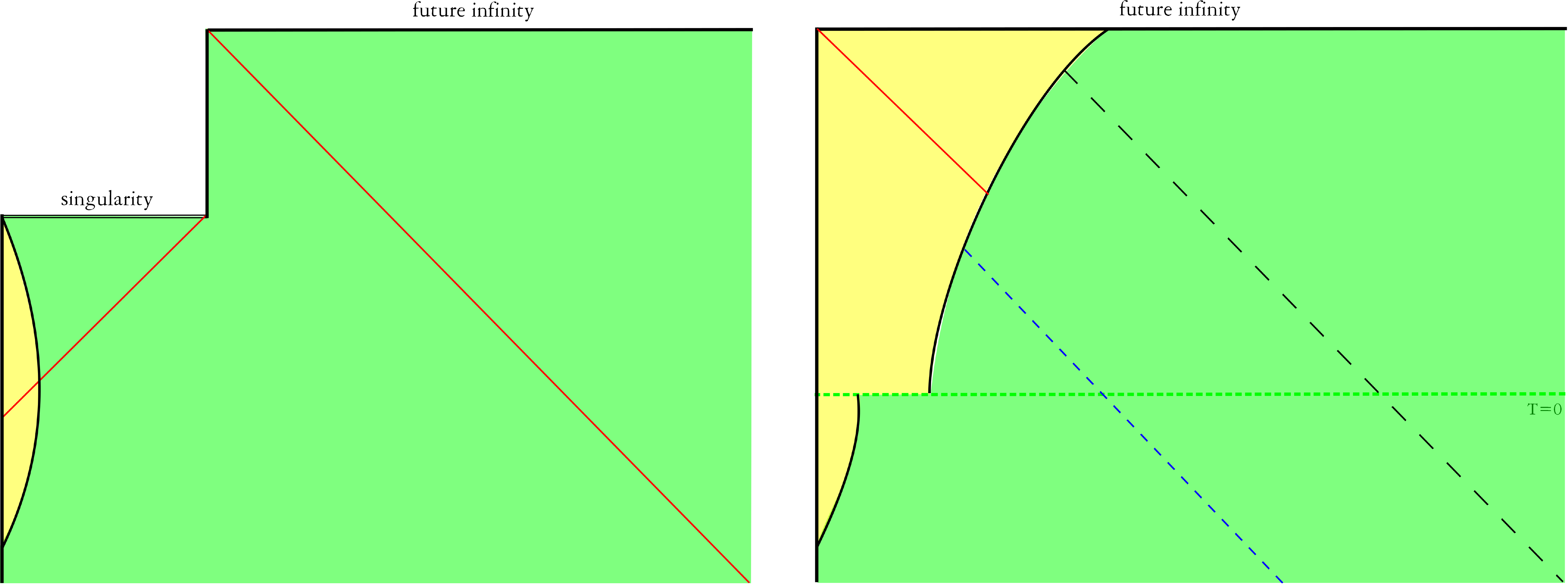}
\caption{\label{fig:dStunnel2}Left: Usual gravitational collapse and evaporation due to Hawking radiation. Right: Tunneling from $\mathrm{dS}_{1}-\mathrm{SdS}_{2}$ to $\mathrm{dS}_{3}-\mathrm{SdS}_{5}$. Black dashed curve is for the false vacuum bubble, while blue dashed curve is for the true vacuum bubble.}
\end{center}
\end{figure}
\begin{figure}
\begin{center}
\includegraphics[scale=0.4]{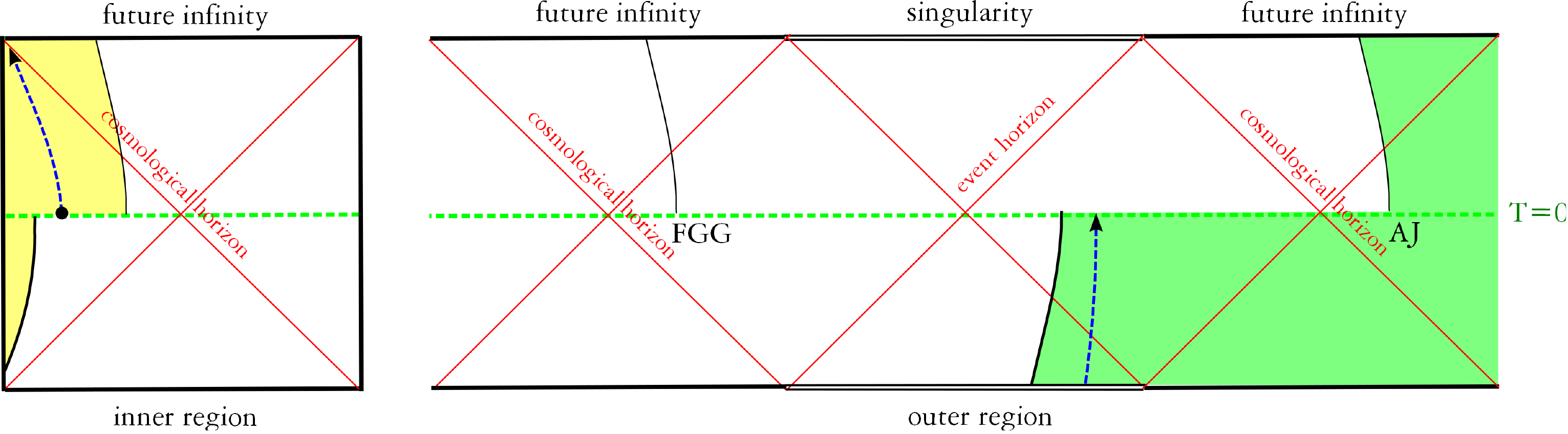}
\caption{\label{fig:dStunnel}Tunneling from $\mathrm{dS}_{1}-\mathrm{SdS}_{2}$ to $\mathrm{dS}_{3}-\mathrm{SdS}_{6}$ (AJ) and $\mathrm{dS}_{3}-\mathrm{SdS}'_{6}$ (FGG). The blue dashed arrow is a typical world-line of an observer.}
\end{center}
\end{figure}

By analyzing the equation of motion, we can classify the shell dynamics as well as possible tunneling channels \cite{Aguirre:2005nt} (Fig.~\ref{fig:signs_dS}). In terms of tunneling processes toward a trivial geometry, we illustrate two important processes.
\begin{itemize}
\item[(i)] \textit{Disappearance of a black hole.} A collapsing shell can tunnel to a bouncing shell with out-going energy \cite{Gregory:2013hja}. This was investigated by Gregory-Moss-Withers \cite{Gregory:2013hja} (Fig.~\ref{fig:dStunnel2}).
\item[(ii)] \textit{Sweep of a bubble.} For some parameter spaces, a tunneling process such as $\mathrm{dS}_{3}-\mathrm{SdS}_{6}$ or $\mathrm{dS}_{3}-\mathrm{SdS}'_{6}$ is possible (Fig.~\ref{fig:dStunnel}). The latter was investigated by Farhi-Guth-Guven (FGG) \cite{Farhi:1989yr} and Fischler-Morgan-Polchinksi \cite{Fischler:1990pk}, while the former was investigated by Aguirre-Johnson (AJ) \cite{Aguirre:2005nt}.
\end{itemize}
The Aguirre-Johnson tunneling can change to a trivial geometry at least for a local observer (blue dashed arrow in Fig.~\ref{fig:dStunnel}). However, in this case, a more correct expression is that the shell sweeps the observer and hence the observer suddenly locates inside the bubble. Hence, \textit{the singularity should be located outside the causal patch of the observer}. Only if we restrict the path integral inside the observer-dependent causal patch, we can say that this tunneling induces a trivial geomtery.

\section{Future perspectives}

Non-perturbative effects will shed some lights on the information loss problem and instantons will help to understand this. We have found various thin-shell instantons, but we need to generalize further: we need to find instantons that result trivial geometries (i) without relying on vacuum structures and (ii) without relying on the thin-shell approximation.

For de Sitter cases, there is an interesting process, so-called the Aguirre-Johnson tunneling. This contributes the cosmological horizon scale. Can this mean that a unitary observer should see a superposition of states or uncertainty around the cosmological horizon? In any case, unitarity of the de Sitter space is less clear in the literature. This may cause subtle issues and we need further investigations.

\newpage

\appendix{Supplementary calculations}

One can simplify the junction equation:
\begin{eqnarray}\label{eq:form}
\left(\frac{dz}{d\eta}\right)^{2} + U(z) &=& - \mathcal{E},\\
U(z) &=& - \left( z^{2} + \frac{\mathcal{C}}{z} + \frac{1}{z^{4}} \right),\\
\mathcal{C} &=& \frac{2 (-L_{-}^{-2} + L_{+}^{-2} + 1)}{\sqrt{(-L_{-}^{-2} + L_{+}^{-2} - 1)^{2} + 4 L_{+}^{-2}}},\\
\mathcal{E}^{3} &=& \frac{1}{\tilde{M}_{+}^{2} \pi^{4} \sigma^{4}} \frac{1}{\left( (-L_{-}^{-2} + L_{+}^{-2} - 1)^{2} + 4 L_{+}^{-2} \right)^{2}},
\end{eqnarray}
where we define parameters by 
\begin{eqnarray}
z &\equiv& \left( \frac{\sqrt{\mathcal{B}^{2} + 64\pi^{2}\sigma^{2}/\ell_{+}^{2}}}{2 M_{+}} \right)^{1/3} r,\\
\eta &\equiv& \left( \frac{\sqrt{\mathcal{B}^{2} + 64\pi^{2}\sigma^{2}/\ell_{+}^{2}}}{8\pi\sigma} \right) t,\\
L_{\pm} &\equiv& 4 \pi\sigma \ell_{\pm},\\
\tilde{M}_{\pm} &\equiv& \frac{M_{\pm}}{\pi \sigma}.
\end{eqnarray}
Now the equation of motion is effectively independent of $\sigma$.

\begin{figure}
\begin{center}
\includegraphics[scale=0.4]{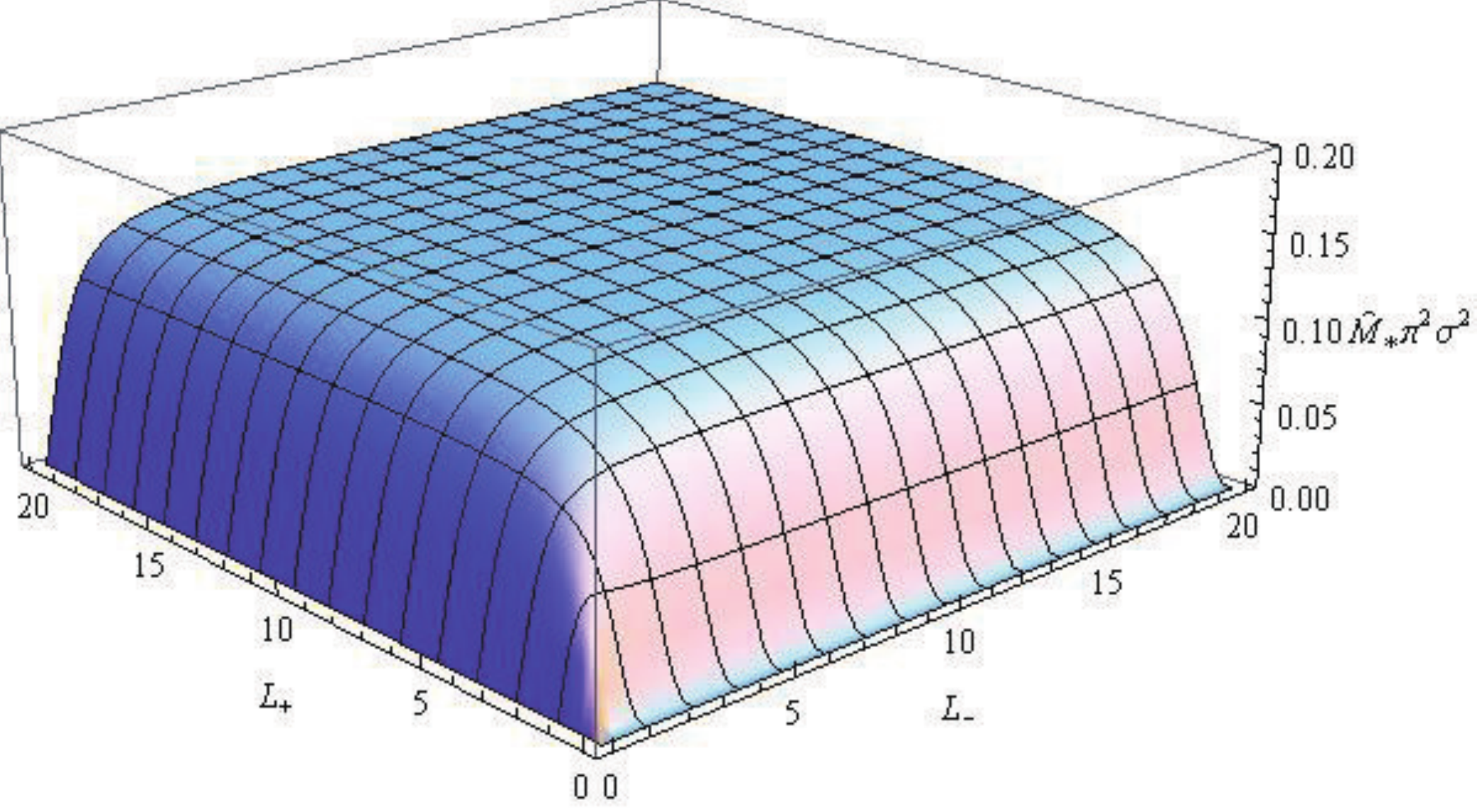}
\caption{\label{fig:mstar}$M_{*} \pi\sigma$ as a function of $L_{\pm}$.}
\end{center}
\end{figure}

In order to obtain symmetric solutions, we need to require the condition: for a $z_{0}$ with $U'(z_{0}) = 0$, there is $\mathcal{E}$ such that $\left| U(z_{0}) \right| < \mathcal{E}$. Hence, $z_{0}$ should satisfy $z_{0}^{6} - \mathcal{C} z_{0}^{3}/2 - 2 = 0$ and we obtain $z_{0}^{3} = (\mathcal{C} + \sqrt{\mathcal{C}^{2} + 32})/4$ to choose $z_{0} > 0$. From this, the existence of symmetric solutions is guaranteed if $M_{+} \leq M_{*}$ (Fig.~\ref{fig:mstar}), where
\begin{eqnarray}
M_{*} \equiv \frac{1}{\pi \sigma}  \frac{\left( (\mathcal{C} + \sqrt{\mathcal{C}^{2} + 32})/4 \right)^{2}}{(-L_{-}^{-2} + L_{+}^{-2} - 1)^{2} + 4 L_{+}^{-2}} \left[3+ \frac{3\mathcal{C}}{2} \left(\frac{\mathcal{C} + \sqrt{\mathcal{C}^{2} + 32}}{4}\right)\right]^{-3/2}.
\end{eqnarray}

After we classify the classical trajectories, we determine signs of $\epsilon$ parameters by comparing extrinsic curvatures:
\begin{eqnarray}\label{eq:ec1}
\beta_{\pm}(r) \equiv \frac{f_{-}(r)-f_{+}(r)\mp 16\pi^{2} \sigma^{2} r^{2}}{8 \pi \sigma r} = \epsilon_{\pm} \sqrt{\dot{r}^{2}+f_{\pm}(r)}. \label{eq:ec2}
\end{eqnarray}
Signs of them are determined by $\beta_{\pm} \propto z^{-3} + \mathcal{D_{\pm}}$, where
\begin{eqnarray}
\mathcal{D_{\pm}} \equiv \frac{-L_{-}^{-2}+L_{+}^{-2}\mp 1}{\sqrt{\left(-L_{-}^{-2}+L_{+}^{-2} - 1 \right)^{2} + 4 L_{+}^{-2}}}.
\end{eqnarray}

By using this, we can understand the asymptotic behaviors of extrinsic curvatures and we can eventually classify possible shell trajectories.
\begin{itemize}
\item[$-$] \textit{Asymptotic behaviors:}
\begin{itemize}
\item[$-$] For collapsing shells, as $r \rightarrow 0$, $\beta_{\pm} \rightarrow +$, and hence bends toward left.
\item[$-$] For expanding shells, as $r \rightarrow \infty$, $\beta_{\pm}$ are up to the signs of $\mathcal{D}_{\pm}$. For false vacuum bubbles, $\epsilon_{+}$ is always negative. For true vacuum bubbles, $\epsilon_{-}$ is always positive.
\end{itemize}
\item[$-$] \textit{Change of signs of extrinsic curvatures:}
\begin{itemize}
\item[$-$] $0 \leq \mathcal{D}_{+} \leq \mathcal{D}_{-}$: $\beta_{\pm}$ do not change the sign.
\item[$-$] $\mathcal{D}_{+} \leq 0 \leq \mathcal{D}_{-}$: $\beta_{+}$ can change the sign.
\item[$-$] $\mathcal{D}_{+} \leq \mathcal{D}_{-} \leq 0$: both of $\beta_{+}$ and $\beta_{-}$ can change the sign.
\end{itemize}
\end{itemize}

We listed all possible trajectories of symmetric solutions in Fig.~\ref{fig:signs_dS}. However, for consistency, only some of them will be allowed in reality.

First, we consider the collapsing solutions. Due to the asymptotic behavior, $\mathrm{dS}_{2}$, $\mathrm{SdS}_{1}$, and $\mathrm{SdS}_{4}$ are not allowed. Therefore, possible solutions are $\mathrm{dS}_{1}-\mathrm{SdS}_{2}$ and $\mathrm{dS}_{1}-\mathrm{SdS}_{3}$.

Second, we consider the bouncing solutions. One thing we should keep in mind is that $\epsilon_{+}=1$ and $\epsilon_{-}=-1$ cannot be true at the same time, as long as the tension is positive. If the shell does not change the sign of extrinsic curvatures, then $\mathrm{dS}_{3}-\mathrm{SdS}_{5}$, $\mathrm{dS}_{3}-\mathrm{SdS}_{6}$, $\mathrm{dS}_{4}-\mathrm{SdS}_{6}$ and their counterparts $\mathrm{SdS}'$ are possible.

If the shell changes the sign of extrinsic curvatures, one more thing we should keep in mind is that whenever $\beta_{+}$ or $\beta_{-}$ changes the sign, it should be negative in the $r \rightarrow \infty$ limit. Then the allowed solutions are $\mathrm{dS}_{3}-\mathrm{SdS}_{7}$, $\mathrm{dS}_{5}-\mathrm{SdS}_{6}$, $\mathrm{dS}_{5}-\mathrm{SdS}_{7}$, and their counterparts $\mathrm{SdS}'$.

Especially, for true vacuum bouncing solutions, only $\mathrm{dS}_{3}$ is allowed, since $\epsilon_{-} = +1$ in the $r \rightarrow \infty$ limit. For false vacuum bouncing solutions, only $\mathrm{SdS}_{6}$ and $\mathrm{SdS}_{7}$ are allowed, since $\epsilon_{+} = -1$ in the $r \rightarrow \infty$ limit. One more interesting limit is the large tension limit (fixing $L_{\pm}$): $\mathcal{D}_{+} < 0$ and $\mathcal{D}_{-} > 0$. Hence, this emerges to only $\mathrm{dS}_{3}-\mathrm{SdS}_{6}$.

\bibliographystyle{ws-procs961x669}

\newpage

\section*{Acknowledgment}

The author would like to thank Pisin Chen, Guillem Dom\'enech, Yao-Chieh Hu, Bum-Hoon Lee, Wonwoo Lee, Yen Chin Ong, Don N. Page, and Misao Sasaki for comments and stimulated discussions. This work is supported by Leung Center for Cosmology and Particle Astrophysics (LeCosPA) of National Taiwan University (103R4000).

\end{document}